\documentclass[%
 aip,
 amsmath,amssymb,
 reprint,%
]{revtex4-1}

\usepackage{graphicx}

\begin{document}

\preprint{AIP/123-QED}

\title{Development of a Ferromagnetic Resonance Measurement System Using NanoVNA}

\author{Reo Fukunaga}
\affiliation{Department of Material Science, Graduate School of Science, University of Hyogo, Ako, Hyogo 678-1297, Japan}

\author{Ryunosuke Takahashi}
\affiliation{Department of Material Science, Graduate School of Science, University of Hyogo, Ako, Hyogo 678-1297, Japan}

\author{Tetsuro Ueno}
\affiliation{Synchrotron Radiation Research Center, Kansai Institute for Photon Science, National Institutes for Quantum Science and Technology, Sayo,
Hyogo 679-5148 Japan}
\affiliation{Quantum Materials and Applications Research Center, Takasaki Institute for Advanced Quantum Science, National Institutes for Quantum Science and Technology, Takasaki 370-1292, Japan}

\author{Hiroki Shoji}
\author{Yoshihiko Togawa}
\affiliation{Department of Physics, Osaka Metropolitan University, Sakai, Osaka 599-8531 Japan}

\author{Hiroki Wadati}
\email{wadati@sci.u-hyogo.ac.jp}
\affiliation{Department of Material Science, Graduate School of Science, University of Hyogo, Ako, Hyogo 678-1297, Japan}
\affiliation{Institute of Laser Engineering, The University of Osaka, Suita, Osaka 565-0871, Japan}

\begin{abstract}
Ferromagnetic resonance (FMR) is a fundamental technique for probing magnetization dynamics in spintronic and magnetic materials. However, conventional FMR measurements rely on broadband vector network analyzers (VNAs), whose high cost limits accessibility for small laboratories and educational environments. To overcome this barrier, we have developed a compact and low-cost FMR measurement platform—the NanoVNA-FMR system—based on a commercially available NanoVNA. The setup integrates an electromagnet and a coplanar waveguide (CPW) and is fully automated using Python scripts. This enables synchronized magnetic-field sweeping, S-parameter acquisition, and real-time visualization. The system successfully captures clear FMR spectra that exhibit systematic shifts in resonance frequency with increasing magnetic field. The results are in excellent agreement with those obtained using a conventional VNA-based FMR system, confirming the quantitative reliability of the NanoVNA approach. Additionally, a 3D-printed sample holder further reduces overall system cost. These results demonstrate that the NanoVNA-FMR system provides a practical, accurate, and accessible alternative for quantitative magnetic characterization and educational applications.
\end{abstract}

\maketitle

\section{Introduction}
Ferromagnetic resonance (FMR) is a fundamental technique for evaluating dynamic magnetic properties such as magnetic anisotropy and saturation magnetization.\cite{kittel1948fmr} Frequency-, field-, and time-domain implementations of FMR have been systematically compared, \cite{neudecker2006comparison} and anomalous high-frequency responses in ferromagnets were already recognized in early work by Griffiths. \cite{griffiths1946anomalous} In magnetic insulators such as yttrium iron garnet (YIG), FMR is widely used to determine damping, effective magnetization, and spectroscopic $g$-factors, and has established YIG as a model material for magnonics and spintronics.\cite{hauser2016recrystallization,cheshire2022absence,lee2016lowfreq,onbasli2014yig,emori2015pseudomorphic} Modern broadband FMR setups typically employ vector network analyzers (VNAs) to measure the transmission coefficient $S_{21}$ in the GHz range. Early broadband coplanar waveguide (CPW)-FMR implementations were demonstrated by Montoya \textit{et al}. for ultrathin magnetic films in 2014,\cite{montoya2014broadband} and the sensitivity of VNA-based spectrometers was later enhanced by the field-differential detection method introduced by Tamaru \textit{et al}. in 2018.\cite{tamaru2018vna} More recently, Montanheiro \textit{et al}. achieved real-time visualization of Lorentzian resonance curves by synchronizing a VNA with an Arduino-controlled magnetic-field sweep.\cite{montanheiro2022realtime} The latest development in this line of research is the OpenFMR project,\cite{schneider2025openfmr}, which provides a fully
Python-automated, open-source broadband FMR platform. Despite these technological advances, the cost and operational complexity of broadband VNAs continue to limit their accessibility for small laboratories and educational environments.

Recently, compact and inexpensive portable VNAs such as the NanoVNA have attracted attention. Although their dynamic range and calibration accuracy are inferior to those of high-end VNAs, they support basic S-parameter measurements and can be controlled via USB using Python. Such devices have already been applied to scientific instrumentation, for example, in the development of a high-frequency dielectric spectrometer using a portable VNA,\cite{erkoreka2024dielectric} but, to our knowledge, there have been no peer-reviewed reports of FMR measurements based on NanoVNA hardware.

In this work, we develop a NanoVNA-based FMR measurement system that integrates a CPW, a DC electromagnet, and fully automated Python-based control. We demonstrate clear FMR spectra from a Y$_3$Fe$_5$O$_{12}$ (YIG) thin film on a Gd$_3$Ga$_5$O$_{12}$ (GGG) substrate and show that the extracted resonance behavior agrees quantitatively with measurements obtained using a conventional broadband VNA, establishing the NanoVNA as a practical low-cost platform for quantitative magnetic spectroscopy.

\section{Experimental setup}
In this study, FMR measurements were performed with the external magnetic field applied in-plane, parallel to the surface of the YIG thin film. 
This geometry enables practical observation of resonance behavior under in-plane magnetization conditions, which is essential for accurately evaluating the intrinsic FMR properties of the sample.

The measurement system was built around the NanoVNA (NanoVNA-F V2), a compact, low-cost VNA. The NanoVNA was connected to a personal computer via USB and controlled using custom Python scripts. During measurements, high-frequency signals generated by the NanoVNA were transmitted through a coaxial cable to a CPW, on which the magnetic thin-film sample was placed. The transmitted signal ($S_{21}$) was subsequently measured by the NanoVNA, and the absorption peak under resonance conditions was extracted from the frequency-dependent transmission spectra. An overview of the complete measurement setup is shown in Fig.~1.

\subsection{High-Frequency Measurement}
In this study, a compact, low-cost NanoVNA was used to measure S-parameters. The NanoVNA supports high-frequency signals up to 3~GHz and enables acquisition of the transmission coefficient $S_{21}$ as a function of frequency.

The measurement configuration consisted of outputting a high-frequency signal from P1, transmitting it through the sample, and receiving it at P2. Both ports were connected to a CPW using high-performance SMA coaxial cables (Thorlabs SMM-series), which offer low reflection and insertion loss, ensuring stable signal transmission at GHz frequencies.

During measurement, the frequency range was fixed while the external magnetic field was varied. 
Multiple $S_{21}$ spectra were recorded to evaluate the magnetic-field dependence of the resonance peak corresponding to FMR.

To verify the accuracy of the NanoVNA-based measurements, complementary FMR measurements were also conducted using a broadband benchtop VNA (Rohde \& Schwarz ZVL6) capable of frequency sweeps from 9~kHz to 6~GHz with the same CPW.
The spectra obtained from both systems showed excellent agreement, confirming the quantitative reliability of the NanoVNA-FMR setup.

\subsection{Electromagnet}

A DC-driven electromagnet (TESLA TMSP232-
1204015) was used to apply an external magnetic field. 
In this study, a maximum field of approximately 45~mT was generated in the in-plane direction of the YIG thin film. 
The field was supplied by a programmable DC power source (OWON P4305), which provided stable and precise control of the current through the electromagnet via voltage adjustment.

In the present configuration, an applied voltage of 1~V corresponded to a magnetic field of 2.5~mT. 
Based on this calibration, the magnetic field was swept from 0 to 45~mT in 2.5~mT increments (2.5~mT/V). 
The magnetic field was applied parallel to the film plane (in-plane geometry), and the sample position was carefully adjusted to ensure uniformity of the field across the sample area.

\begin{figure}[htbp]
\centering
\includegraphics[width=1.0\linewidth]{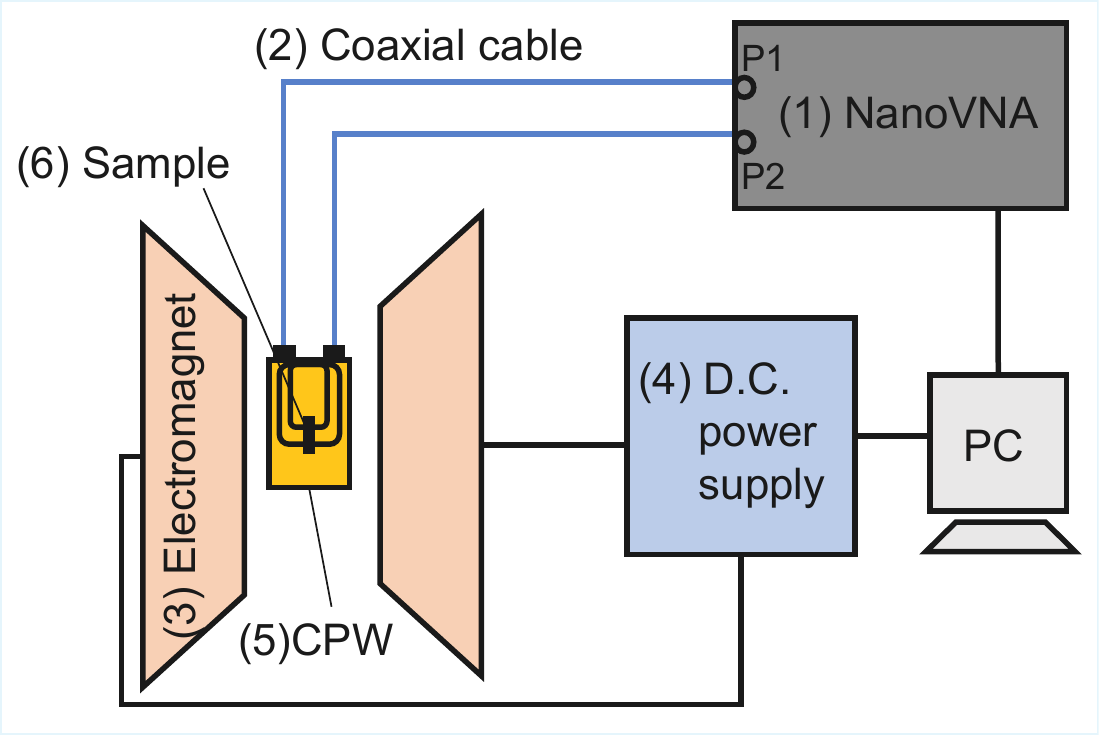}
\includegraphics[width=1.0\linewidth]{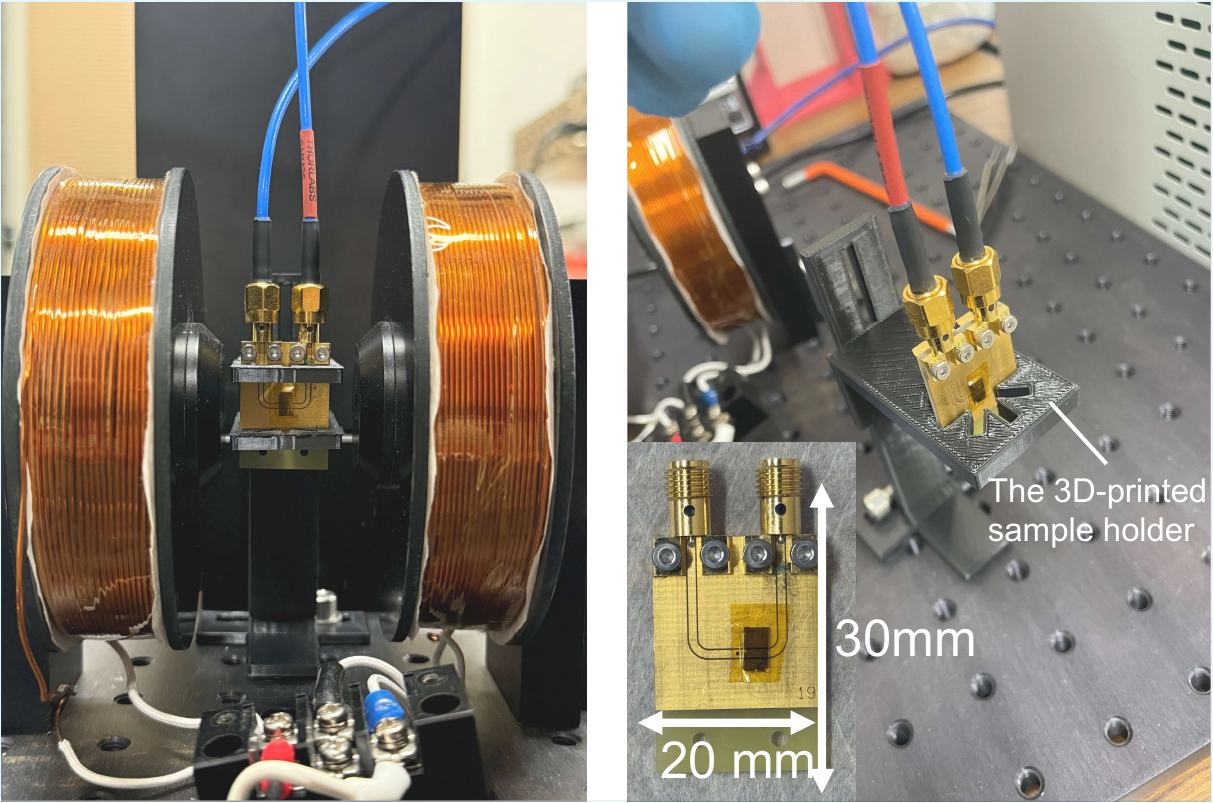}
\caption{(a) Schematic diagram of the NanoVNA-based ferromagnetic resonance (FMR)
measurement setup, showing the electromagnet, coplanar waveguide (CPW), DC power supply
and NanoVNA connections.
(b) Photograph of the CPW and the 3D-printed sample holder used in the
experiment.}
\end{figure}

\subsection{Coplanar Waveguide (CPW)}

A CPW was employed for high-frequency signal transmission and microwave excitation of the magnetic sample. 
The CPW consists of a central signal line flanked by two ground conductors on the same plane. 
High-frequency signals propagating along the signal line generate a localized microwave magnetic field that excites FMR in the thin-film sample placed on top of the CPW.

The CPW used in this study was fabricated by Hayashi Repic Co., Ltd. 
It features a center-conductor width of 1.03~mm and a gap width of 0.19~mm, yielding a characteristic impedance of 50~$\Omega$.

The sample was positioned directly above the central conductor and secured using Kapton tape, which provides excellent heat resistance and electrical insulation. 
To ensure mechanical stability and precise alignment within the magnet gap, a custom holder was fabricated using a 3D printer. 
The holder, made of polylactic acid, was designed to minimize vibration and displacement during measurement. 
A photograph of the CPW and sample holder is shown in Fig.~1(b).

\subsection{Sample}
The sample investigated in this study was a thin film of yttrium iron garnet (Y$_3$Fe$_5$O$_{12}$, YIG) deposited on a (111)-oriented gadolinium gallium garnet (Gd$_3$Ga$_5$O$_{12}$, GGG) substrate. 
The film was grown by pulsed laser deposition (PLD) to a thickness of approximately 2~$\mu$m, with lateral dimensions of 3~mm~$\times$~5~mm.

During measurement, the film side was placed facing upward on the CPW and carefully aligned to fully cover the signal-line region. 

Before conducting the FMR measurements, the static magnetic properties of the film were characterized using a vibrating sample magnetometer to evaluate its in-plane and out-of-plane magnetization behavior, including the saturation magnetization and magnetic anisotropy. The measurements confirmed that the easy axis lies within the film plane, and these results were used to validate the expected resonance conditions under the in-plane magnetic-field geometry.

\vspace{1.5\baselineskip}

\begin{table}[htbp]
\caption{Components used in the NanoVNA-based FMR measurement system.}
\label{tab:components_final}

\renewcommand{\arraystretch}{1.25} 
\setlength{\tabcolsep}{4pt}       

\begin{tabular}{|p{2.6cm}|p{1.6cm}|p{3.4cm}|}
\hline
\textbf{Element} & \textbf{Product name} & \textbf{Performance / Specification} \\
\hline

1) Vector network analyzer &
\href{https://www.amazon.co.jp/dp/B0D8W8969D}{NanoVNA-F V2} &
50~kHz–3~GHz; portable unit with 5000~mAh battery \\
\hline

2) Coaxial cable &
\href{https://www.thorlabs.co.jp/thorproduct.cfm?partnumber=SMM36}{Thorlabs SMM36} &
2.92~mm SMA connectors; DC–36~GHz; low-loss \\
\hline

3) Electromagnet &
\href{https://www.teslanet.co.jp/item/%E5%8D%93%E4%B8%8A%E5%9E%8B%E5%85%89%E5%AD%A6%E7%94%A8%E9%AB%98%E7%A3%81%E5%A0%B4%E7%99%BA%E7%94%9F%E9%9B%BB%E7%A3%81%E7%9F%B3/}{TESLA TMSP232-1204015} &
C-frame electromagnet; field strength up to $\sim$400 ~mT \\
\hline

4) DC power supply &
\href{https://www.amazon.co.jp/dp/B0BCKT4GY5}{OWON SPE6103} &
Programmable DC source; 0–60~V / 3~A output \\
\hline

5) Coplanar waveguide (CPW) &
Hayashi Repic Co., Ltd. &
50 $\Omega$ characteristic impedance; microwave-compatible substrate \\
\hline

\end{tabular}
\end{table}

\section{Measurement Procedure and Software}
In this study, a fully automated measurement system was developed in which both the NanoVNA and the DC power supply were controlled via a custom Python script. 
The program coordinated the magnetic-field sweep and S-parameter acquisition, enabling reproducible and efficient collection of FMR spectra in the frequency domain.

The measurement procedure consisted of the following steps:

\begin{enumerate}
    \item{Sample placement:} The YIG thin-film sample was positioned on the center conductor of the CPW with the film side facing upward. The sample was lightly secured using Kapton tape to prevent displacement during measurement.

    \item{Field initialization:} The external magnetic field was set to 0~mT using the DC electromagnet.

    \item{Frequency setup:} The NanoVNA was initialized and configured to sweep across the frequency range of 50~kHz to 3~GHz.

    \item{Reference spectrum acquisition:} A baseline $S_{21}$ spectrum was recorded at 0~V (corresponding to 0~mT) and saved in decibel (dB) scale as a reference.

    \item{Field sweep:} The applied voltage was incremented from 0~V to 18~V in 0.5~V steps, corresponding to magnetic fields from 0 to 45~mT (2.5~mT per volt). At each field value, an $S_{21}$ spectrum was acquired and stored.

    \item {Analysis:} For each field step, a differential transmission spectrum $\Delta S_{21}$ was calculated relative to the 0~mT reference to highlight field-dependent absorption features.

    \item{Peak extraction:} The resonance frequency was identified from the minimum point of each $S_{21}$ spectrum and recorded along with the corresponding magnetic-field value.

    \item{Visualization and fitting:} The acquired data were used to generate\\
(i) field-dependent $S_{21}$ spectra.\\
(ii) a Kittel plot, defined as the resonance frequency $f$ extracted from
Lorentzian fits plotted as a function of the external magnetic field $\mu_{0}H$,
from which $\gamma$ and $\mu_{0}M_{\mathrm{eff}}$ were determined.\\
(iii) a two-dimensional color map of absorption intensity.

\end{enumerate}

The Python control script included modules for port scanning, frequency configuration, S-parameter acquisition, file management, and real-time visualization, forming a closed-loop automation routine. 
All voltage, frequency, and absorption data were exported as comma-separated value files, enabling subsequent curve fitting and statistical analysis.

During the measurement, the software displayed the $S_{21}$ spectra and the evolution of the resonance peaks in real time using Python’s plotting library (Matplotlib). 
A screenshot of the user interface is shown in Fig.~2, illustrating the automated data acquisition and live visualization of the FMR response.

\begin{figure*}[t]
    
    \centering
    \includegraphics[width=\linewidth]{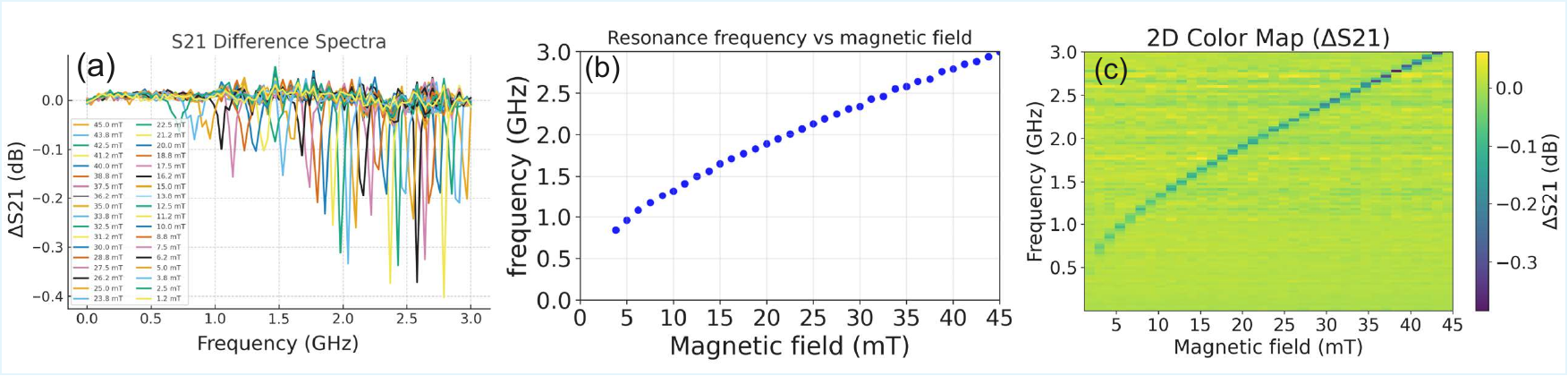}
    \caption{Screenshot of the Python-based control interface used during FMR measurements.
(a) Reference-processed transmission spectra ($S_{21}$), 
(b) automatically extracted resonance peak positions at each field, and 
(c) a color map generated from the measured frequency--field dependence. 
The program performs automatic magnetic-field sweeping and real-time visualization of the transmission spectra acquired by the NanoVNA.
}
    \vspace{-\intextsep} 
\end{figure*}

\section{Results}

FMR measurements were carried out by acquiring the transmission spectra
$S_{21}$ at each applied magnetic field and computing the differential
spectra relative to the 0~mT reference. In this work, the complex transmission coefficient measured by the NanoVNA is denoted as
$S_{21}^{\mathrm{lin}}(f,H)$, and its magnitude in decibels is defined as
\begin{equation}
   S_{21}^{\mathrm{dB}}(f,H)
   = 20 \log_{10} \left| S_{21}^{\mathrm{lin}}(f,H) \right|.
\end{equation}
The differential transmission spectra are then given by
\begin{equation}
   \Delta S_{21}^{\mathrm{dB}}(f,H)
   = S_{21}^{\mathrm{dB}}(f,H)
   - S_{21}^{\mathrm{dB}}(f,0~\mathrm{mT}),
   \label{eq:deltaS21}
\end{equation}
where $S_{21}^{\mathrm{dB}}(f,0~\mathrm{mT})$ represents the reference
spectrum measured at zero magnetic field.
Representative differential transmission spectra
$\Delta S_{21}^{\mathrm{dB}}(f,H)$ at selected field values are shown in
Fig.~3.

To quantitatively evaluate the resonance behavior, each spectrum was fitted using a Lorentzian function. 
This fitting procedure allowed an accurate determination of the resonance frequency and linewidth for each magnetic field value. 
In all cases, the spectra showed clear resonance peaks with sufficient signal-to-noise ratio, confirming the robustness of the NanoVNA-based measurements.

Fig.~4. presents a two-dimensional color map constructed from the acquired $S_{21}$ spectra, with frequency plotted along the horizontal axis and magnetic field along the vertical axis. 
The color scale represents the change in absorption intensity (in dB). 
The resonance band appears as a well-defined, high-absorption region that shifts continuously with increasing field, providing a clear visualization of the FMR dispersion relation.

The resonance frequency $f$ for each magnetic field was quantitatively determined
by fitting the differential transmission spectra with a Lorentzian function,
rather than simply taking the minimum point of the $S_{21}$ curve.
The extracted resonance frequencies and their dependence on the magnetic field
strength $\mu_{0}H$ are shown in Fig.~5.
The field dependence of $f$ follows the well-known Kittel equation:

\begin{equation}
f = \frac{\gamma}{2\pi} \sqrt{\mu_0 H \left( \mu_0 H + \mu_0 M_{\text{eff}} \right)}
\label{eq:kittel_inplane}
\end{equation}

where $\gamma$ is the gyromagnetic ratio and $\mu_0 M_{\text{eff}}$ is the effective magnetization. By fitting the experimental data to this relation, we obtained $\gamma/2\pi$ = 28.06~GHz/T, which agrees well with the literature value of 28.18~GHz/T for high-quality YIG thin films~\cite{lee2016lowfreq}.

\begin{figure}
\centering
\hspace*{0.6cm}
\includegraphics[width=1.0\linewidth]{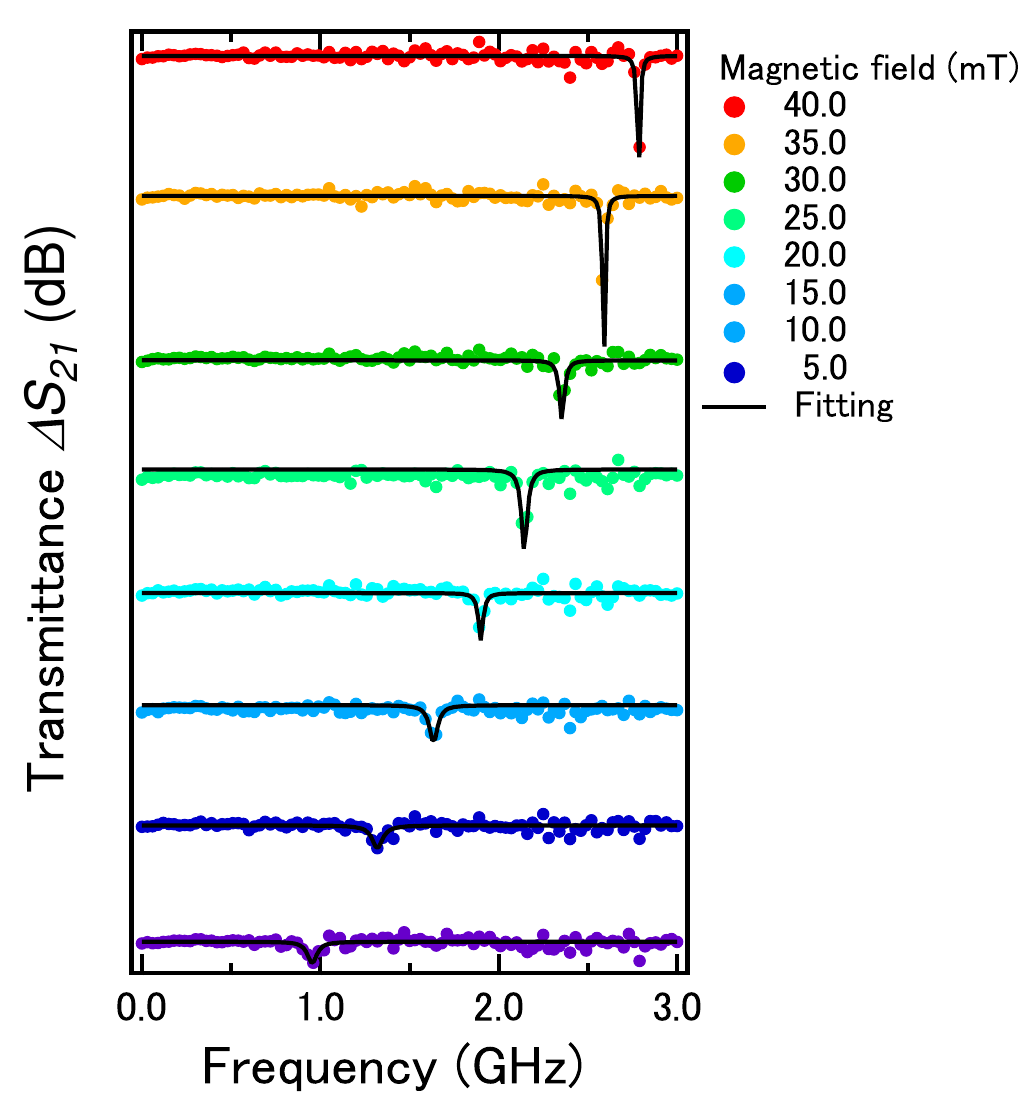}
\caption{Representative differential transmission spectra ($ S_{21}$) measured at magnetic fields ranging from 5.0 to 40~mT in steps of 5~mT. Each spectrum exhibits a clear resonance peak that shifts toward higher frequency with increasing field strength, indicating field-dependent ferromagnetic resonance behavior. All spectra are fitted using a Lorentzian function.}
\end{figure}

To assess the performance of the NanoVNA-based system, comparative measurements were performed using a benchtop VNA capable of frequency sweeps up to 6~GHz. The results are summarized in Fig.~5, where the left panel displays the NanoVNA data over the 50 kHz–3 GHz range, and the right panel compares these results with those from the benchtop VNA. The resonance peaks, linewidths, and field-dependent frequency shifts are in excellent agreement between the two systems, demonstrating that the NanoVNA provides reliable, quantitative FMR data despite its compact, inexpensive design.

Overall, these results confirm that the NanoVNA-FMR system accurately captures the essential features of FMR in YIG thin films, providing sufficient resolution and reproducibility for both research and educational applications.

\begin{figure}
    \centering
    \includegraphics[width=1.0\linewidth]{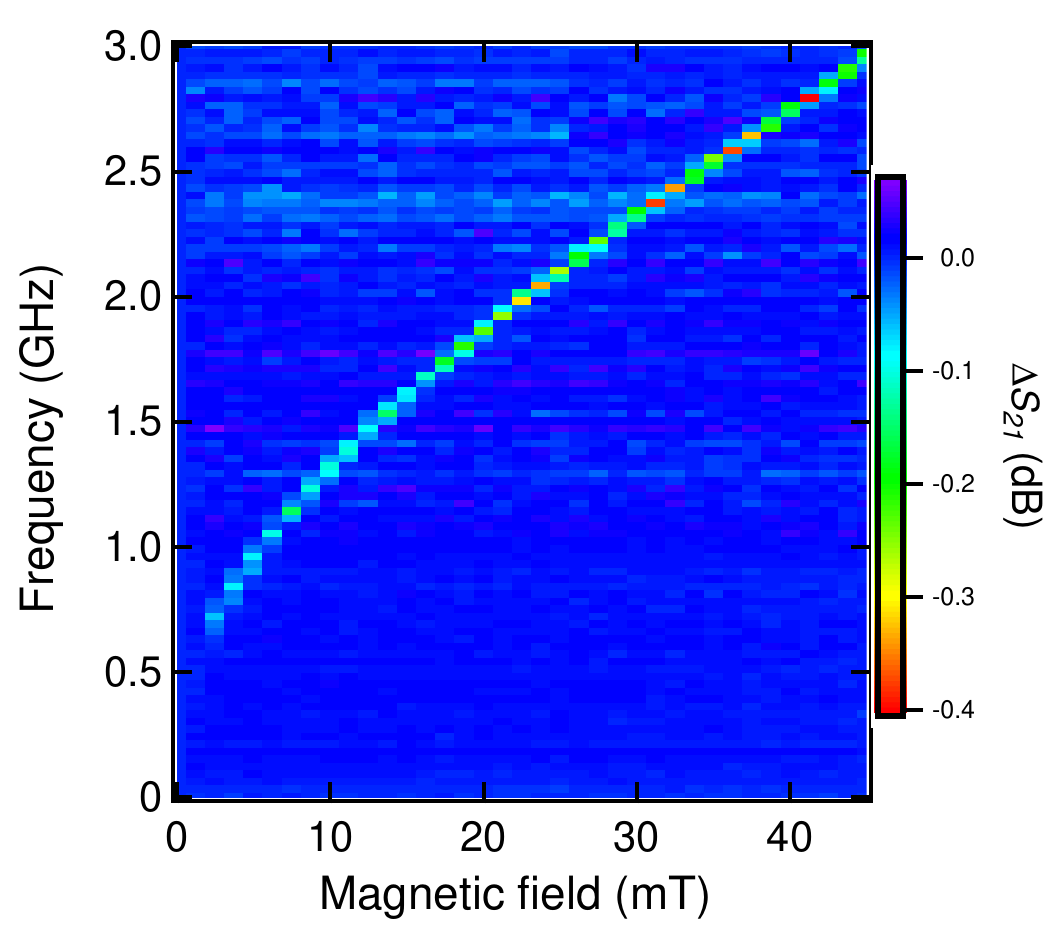}
    \caption{Two-dimensional color map of the FMR absorption intensity as a function of frequency and magnetic field. The resonance band appears as a distinct high-absorption region that shifts continuously with increasing field, providing an intuitive visualization of the FMR dispersion.}
    \label{fig:enter-label}
\end{figure}

\begin{figure}
    \centering
 
    \includegraphics[width=1.0\linewidth]{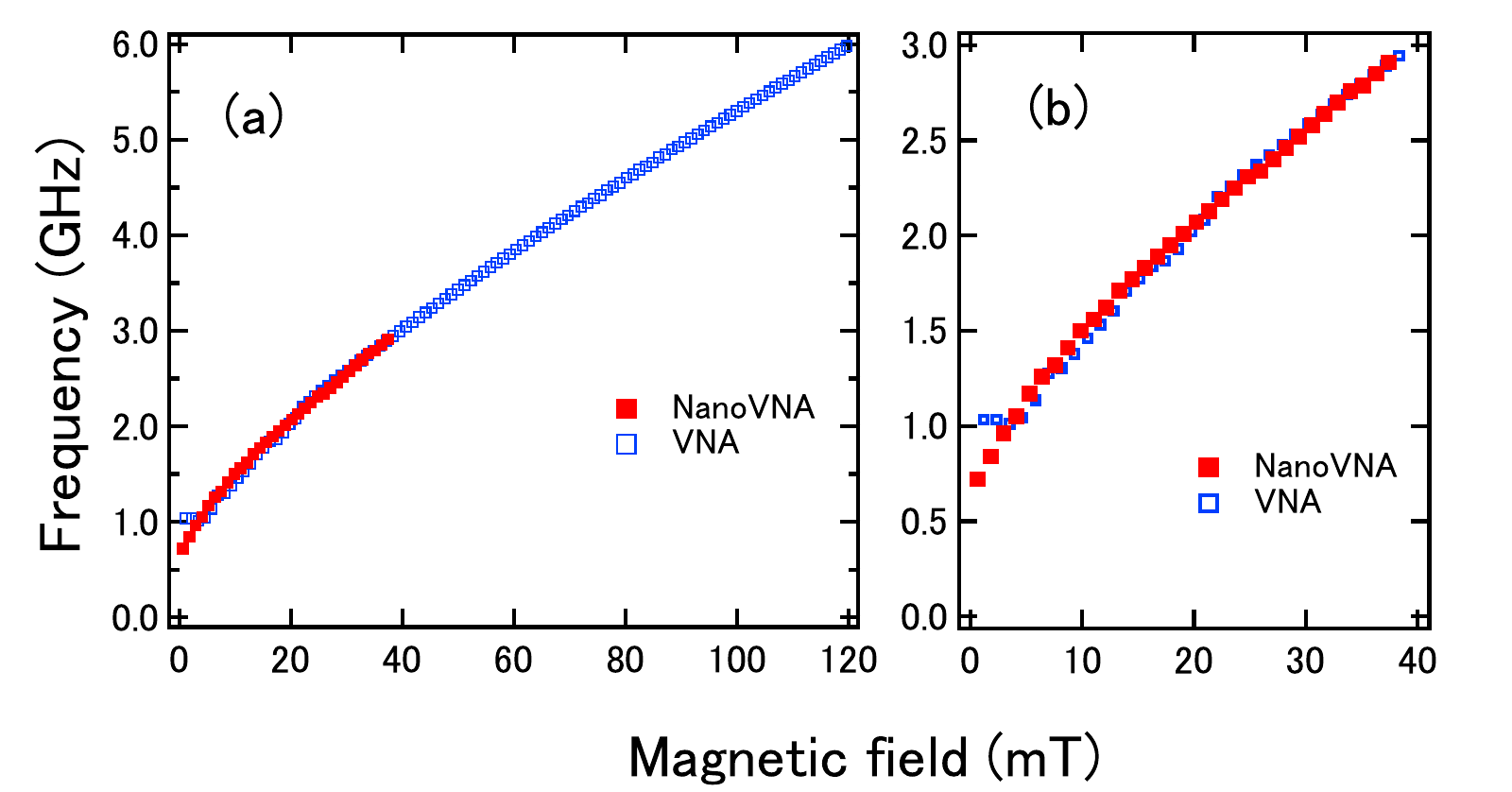}
    \caption{(a) Comparison between the results obtained using the NanoVNA and those measured with a conventional VNA. 
(b) Enlarged view of the measurement range of the NanoVNA (50~kHz–3~GHz).}
    \label{fig:placeholder}
\end{figure}

\section{Discussion}
The results presented above demonstrate that the NanoVNA-based measurement system can successfully capture the fundamental FMR characteristics of YIG thin films. The resonance frequency shifted systematically with the applied magnetic field, and the spectra showed distinct absorption peaks with a sufficiently high signal-to-noise ratio.

From the Kittel fitting, the extracted gyromagnetic ratio, $\gamma/2\pi = 28.06$ GHz/T, and the effective magnetization were consistent with reported values for high-quality YIG thin films, confirming the quantitative accuracy of the developed setup. 

Specifically, the effective magnetization was determined to be 
$\mu_{0} M_{\mathrm{eff}} = 249~\mathrm{mT}$, corresponding to
$4\pi M_{\mathrm{eff}} \approx 2490~\mathrm{G}$ in cgs units. This value is fully
consistent with previous reports on high-quality YIG thin films. For example, Emori
\textit{et al}. observed $4\pi M_{\mathrm{eff}} = 2460$--$2670~\mathrm{G}$ for
pseudomorphic YIG grown on GGG,\cite{emori2015pseudomorphic} while Onbaşlı
\textit{et al}. reported $\mu_{0} M_{\mathrm{s}} \approx 0.17~\mathrm{T}$ for
recrystallized YIG films.\cite{onbasli2014yig} Comparable effective magnetization
values have also been found in amorphous or ultrathin YIG films prepared by PLD.\cite{hauser2016recrystallization, ding2020pma, haidar2023ultrathin}
These comparisons confirm that the extracted $M_{\mathrm{eff}}$ of our YIG film lies well within the expected range.

Furthermore, the comparison with a conventional VNA showed excellent agreement in both the resonance field and the linewidth, indicating that the NanoVNA provides reliable FMR measurements despite its compact, inexpensive design.

It is important to note, however, that the NanoVNA has certain limitations compared to high-end benchtop VNAs. 
The dynamic range and absolute calibration accuracy are lower, which can reduce sensitivity when detecting small linewidth variations or performing broadband measurements above 3~GHz. 
Nevertheless, for relative measurements—such as tracking resonance-field shifts, monitoring magnetic anisotropy, or studying spin-pumping efficiency—the performance of the NanoVNA-FMR system is more than sufficient.

A key advantage of this system is its Python-based automation, which enables high-throughput, reproducible data acquisition with minimal user intervention. This feature significantly reduces human error, shortens measurement time, and allows real-time visualization of magnetic-field-dependent spectra. Such capabilities make the setup well-suited not only for laboratory-scale research but also for educational demonstrations of microwave magnetism.

The broader literature on YIG and garnet films suggests several natural extensions of the present work. Recent experiments have used all-optical FMR in thin YIG films to probe local micromagnetic parameters,\cite{schmoranzerova2023thermal}, and low-temperature FMR has been employed to extract temperature-dependent anisotropy constants in garnet films.\cite{panin2025anisotropy} In addition, detailed studies of the YIG $g$-factor and spin mixing\cite{cheshire2022absence} and broadband CPW-based FMR systems for ultrathin films\cite{montoya2014broadband} provide important benchmarks for future refinements of NanoVNA-based instrumentation.

Overall, the present work establishes the NanoVNA as a practical and accessible platform for quantitative magnetic spectroscopy. With minor hardware and software extensions, the system could be adapted for advanced studies, including damping-constant extraction, temperature-dependent FMR, and angular-dependent measurements. A NanoVNA-based portable FMR system could also be deployed for fieldwork in rock magnetism or paleomagnetism,\cite{kind2012rockmagnetism}, further expanding its potential as a versatile tool for both materials research and magnetism education.

\section{Conclusion}
In this study, we developed and evaluated a compact, low-cost FMR measurement system based on a commercially available NanoVNA. The system integrates a CPW, an electromagnet, and fully automated Python control, enabling synchronized magnetic-field sweeping and frequency-domain $S_{21}$ measurements.

Clear FMR spectra were obtained from YIG thin films, showing systematic shifts in resonance frequency with increasing magnetic field that followed the Kittel relation. Comparative measurements using a conventional broadband VNA demonstrated excellent agreement in both resonance field and linewidth, confirming the quantitative reliability of the NanoVNA-based setup.

The developed NanoVNA-FMR system significantly reduces the instrumentation cost while maintaining sufficient precision for quantitative magnetic analysis. Its compactness, ease of use, and automation make it particularly suitable for educational environments, small-scale laboratories, fieldwork, and initial research on spintronic or magnonic materials.

Future improvements could include extending the frequency range, enhancing field strength, and incorporating additional analysis modules for damping-constant extraction and angular-dependent FMR measurements. Overall, the NanoVNA-FMR platform provides a practical and accessible foundation for advancing both experimental magnetism education and research.

\section*{Acknowledgments}
This work was supported by the JSPS KAKENHI under Grant Nos. 23K25805 and 25H01251, and the MEXT Quantum Leap Flagship Program (MEXT Q-LEAP) under Grant No. JPMXS0118068681. TU also acknowledges the support of the Murata Science and Education Foundation.

\section*{AUTHOR DECLARATIONS}
\subsection*{Conflict of Interest}
The authors have no conflicts to disclose.
\subsection*{Author Contributions}
Reo Fukunaga: Investigation (equal); Methodology (equal);
Writing - original draft (equal).
Ryunosuke Takahashi: Investigation (equal); Methodology (equal); Writing – review \& editing (equal).
Tetsuro Ueno: Investigation (equal); Methodology (equal); Writing – review \& editing (equal).
Hiroki Shoji: Investigation (equal); Methodology (equal); Writing – review \& editing (equal).
Yoshihiko Togawa: Investigation (equal); Methodology (equal); Writing – review \& editing (equal).
Hiroki Wadati: Conceptualization (equal); Funding acquisition (equal); Investigation (equal); Methodology (equal); Writing – review \& editing (equal).\\

\section*{Data Availability Statement}
The data that support the findings of this study are available
from the corresponding author upon reasonable request. The Python file used in the measurements is available at https://github.com/hwadati/NanoVNA.

\end{document}